\pdfoutput=1
\documentclass{article}
\usepackage{spconf}
\usepackage{amsmath,graphicx,xcolor}
\usepackage{booktabs}
\usepackage{url}
\usepackage{microtype}
\usepackage{hyperref}
\usepackage{fancyhdr}
\usepackage{nopageno}

\fancypagestyle{firstpage}{%
  \pagenumbering{gobble}
  \lfoot{\footnotesize © 2021 IEEE.  Personal use of this material is permitted.  Permission from IEEE must be obtained for all other uses, in any current or future media, including reprinting/republishing this material for advertising or promotional purposes, creating new collective works, for resale or redistribution to servers or lists, or reuse of any copyrighted component of this work in other works.}
}

\title{A Hybrid CNN-BiLSTM Voice Activity Detector}
\name{Nicholas Wilkinson, Thomas Niesler}
\address{Department of Electrical and Electronic Engineering, Stellenbosch University, South Africa\\
nwilkinson@sun.ac.za, trn@sun.ac.za}
\begin{document}

\maketitle
\thispagestyle{firstpage}
\newpage

\begin{abstract}
This paper presents a new hybrid architecture for voice activity detection (VAD) incorporating both convolutional neural network (CNN) and bidirectional long short-term memory (BiLSTM) layers trained in an end-to-end manner.
In addition, we focus specifically on optimising the computational efficiency of our architecture in order to deliver robust performance in difficult \textit{in-the-wild} noise conditions in a severely under-resourced setting.
Nested k-fold cross-validation was used to explore the hyperparameter space, and the trade-off between optimal parameters and model size is discussed.
The performance effect of a BiLSTM layer compared to a unidirectional LSTM layer was also considered.
We compare our systems with three established baselines on the AVA-Speech dataset.
We find that significantly smaller models with near optimal parameters perform on par with larger models trained with optimal parameters.
BiLSTM layers were shown to improve accuracy over unidirectional layers by $\approx$2\% absolute on average.
With an area under the curve (AUC) of 0.951, our system outperforms all baselines, including a much larger ResNet system, particularly in difficult noise conditions.
\end{abstract}



\section{Introduction}
\label{sec:intro}

\vspace{-2mm}

Voice activity detection (VAD) is the task of identifying speech and non-speech portions within an audio signal.
Real world speech signals are often noisy and occur within portions of extended silence, environmental noise or music.
A VAD, therefore, is an important preprocessing step for many real world speech processing systems, speech enhancement, speaker identification and automatic speech recognition (ASR).

Efficient, accurate VAD has been a topic of research interest since the 1960s.
Early systems applied a threshold to the energy of the signal to detect the presence of speech \cite{TASI}.
These first systems were improved upon by applying adaptive thresholds to a number of temporal and spectral features \cite{G.729, GSM06.32}, and introducing a hangover period to avoid truncation of speech.

When the signal-to-noise (SNR) ratio is high, these simple systems deliver satisfactory performance.
However, as the SNR decreases, their performance degrades considerably.
Statistical model-based approaches have been proposed to address the problem of robust VAD in low SNR environments.
Sohn et. al. \cite{Sohn} consider the discrete Fourier transform (DFT) coefficients of speech and noise to be asymptotically independent Gaussian random variables.
A likelihood ratio test is then applied to identify speech, and a statistical hidden Markov model (HMM) hangover scheme is introduced.
A number of similar statistical approaches have been built on this work, making use of different features \cite{KLTLaplaceGauss}, distributions \cite{Gamma, Multiple_models}, or decision rules \cite{MO_LRT}.
While these statistical model-based schemes can deliver good performance, they fail when presented with difficult non-speech noise, such as music.

Recently, state-of-the-art VAD performance has been achieved through systems based on machine learning.
By treating VAD as a frame-based classification problem, various classifiers can be trained to identify speech/non-speech frames.
Support vector machine (SVM) classifiers have been extensively used to this purpose \cite{SMV1, SVM2, SVM3, SVM4}, and more recently various neural network architectures \cite{DBN_VAD, DNN_VAD1, RNN_VAD, DNN_VAD2}.

Our work is part of a broader project on ASR in a severely resource constrained setting \cite{Biswas2019}.
As such, our goal is to develop a lightweight VAD that is computationally efficient enough to run on a mobile device, yet sufficiently accurate to provide downstream ASR or keyword spotting systems with accurate speech labels.
In previous work, we introduced a VAD based on a convolutional neural network (CNN) classifier followed by Gaussian mixture model-hidden Markov model (GMM-HMM) smoothing scheme \cite{wilkinson}.
The promising performance of CNN audio classifiers and VAD systems in the literature motivated this design \cite{BigCNN, SmallCNN}.
However, we found that the lack of temporal modeling caused frame drops/insertions in speech/non-speech segments respectively.
Our solution was to introduce the GMM-HMM smoothing scheme.
In this paper we present a novel convolutional neural network-bidirectional long short-term memory (CNN-BiLSTM) VAD, which removes the need for such smoothing by modelling the temporal context within a single network trained in an end-to-end fashion.
We explore the hyperparameter space, compare unidirectional and bidirectional LSTM layer performance, and evaluate our model against three strong baseline systems.

\vspace{-2mm}

\section{Data Description}
\label{sec:data}

\vspace{-2mm}

Our experiments are conducted using AVA-Speech, a publicly available dataset of movies densely labeled with speech activity \cite{AVASpeech}.
At time of writing, AVA-Speech consisted of 160 segments from movies hosted on YouTube, each 15 minutes in duration, totalling 40 hours of labelled data.
The segments are densely labelled for speech activity using the following mutually exclusive labels: ``NoSpeech'', ``CleanSpeech'', ``Speech+Music'' and ``Speech+Noise''.
Each segment is human-labelled by 3 annotators and the annotations are merged using a frame-level majority vote.

This dataset provides a diverse set of speakers, acoustic conditions and languages.
Furthermore, movie data provides a good approximation of \textit{in-the-wild} broadcast media, as opposed to synthetically corrupted datasets commonly used for VAD development and testing.
Consequently, the dataset contains roughly equal amounts of speech and non-speech data, and most of the speech data is noisy.
Dataset statistics are given in \mbox{Table \ref{tab:ava_stats}}.
The SNR shown is an estimate obtained from a trained time-frequency-masking-based speech enhancement neural network, as the ground truth SNR is unavailable.

\begin{table}
    \resizebox{\columnwidth}{!}{
    \scriptsize
    \centering
    \begin{tabular}{@{\extracolsep{\fill}}lcccc}
        \toprule
        {\textbf{Label}} &
        \begin{tabular}[c]{@{}c@{}}\textbf{Time}\\ \textbf{(\%)}\end{tabular}&
        \begin{tabular}[c]{@{}c@{}}\textbf{Segments}\\ \textbf{(\%)}\end{tabular}&
        \begin{tabular}[c]{@{}c@{}}\textbf{AvgDur}\\ \textbf{(s)}\end{tabular}&
        \begin{tabular}[c]{@{}c@{}}\textbf{SNR}\\ \textbf{(dB)}\end{tabular} \\
        \midrule
        CleanSpeech & 14.55 & 16.68 & 2.97 & 40.8 \\ 
        Speech+Music & 13.46 & 13.33 & 3.43 & 11.7 \\ 
        Speech+Noise & 24.32 & 25.41 & 3.28 & 16.2 \\
        NoSpeech & 47.68 & 44.57 & 3.68 & N/A \\
        \bottomrule
    \end{tabular}
    }
    \caption{\textit{AVA-Speech} dataset statistics \cite{AVASpeech}.}
    \label{tab:ava_stats}
    \vspace{-4mm}
\end{table}

\vspace{-2mm}

\section{Proposed CNN-BiLSTM VAD}
\label{sec:system}

\vspace{-2mm}

We introduce a compact CNN-BiLSTM hybrid model for VAD.
A hybrid convolutional, long short-term memory, deep neural network (CLDNN) model was first introduced in \cite{CLDNN} and found to outperform previous models for speech recognition tasks.
This inspired a CLDNN for VAD presented in \cite{CLDNNVAD}.
This system, however, differs significantly from the model we present, in that it uses raw-waveform features, one-dimensional convolutions, and unidirectional LSTM layers.
Our model is inspired by the state-of-the-art performance of two-dimensional CNN architectures applied to spectrograms for audio classification tasks \cite{BigCNN}, and the capacity of BiLSTM layers to model temporal sequences.

\begin{figure}
\begin{minipage}[b]{1.0\linewidth}
  \centering
  \centerline{\includegraphics[width=.7\linewidth]{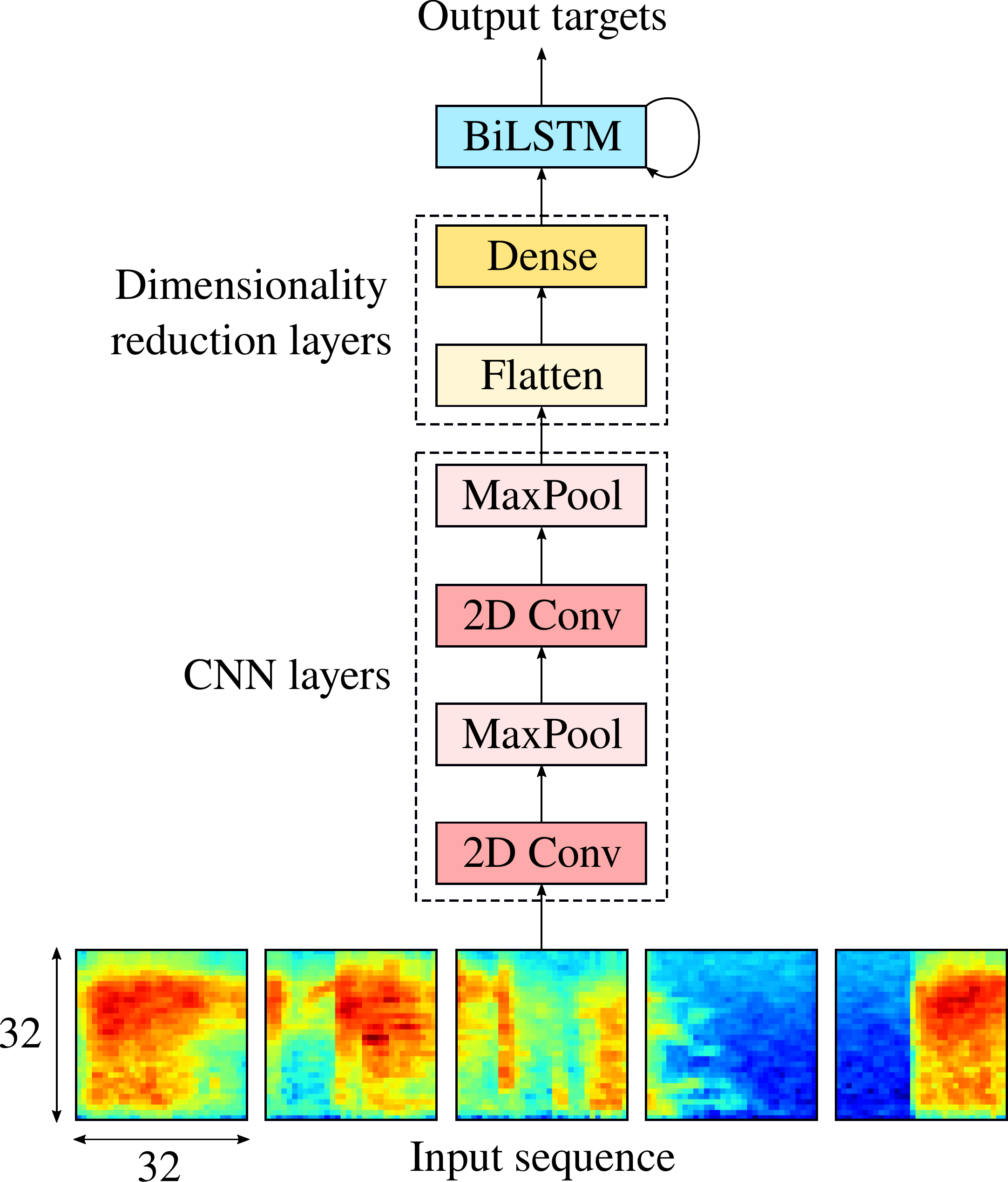}}
\end{minipage}
\vspace{-6mm}
\caption{Block diagram of the CNN-BiLSTM VAD.}
\label{fig:architecture}

\end{figure}

A block diagram of our architecture is shown in \mbox{Figure \ref{fig:architecture}}.
It consists of two two-dimensional convolutional layers, with rectified linear unit (ReLU) activations and max pooling layers.
To further reduce the dimension and therefore the computation, the output from the second max pooling layer is flattened and fed to a dense layer with a ReLU activation.
The embedding from this dense layer is connected to a BiLSTM layer with a tanh activations and sigmoid recurrent activations.
Finally, the BiLSTM is connected to a two-dimensional softmax output, representing speech and non-speech respectively.
This system is implemented in Python, using TensorFlow (v2.0.0) and Keras (v2.2.4-tf).
All models are trained with the Adam optimiser \cite{Adam} and a binary cross-entropy loss function.

\vspace{-2mm}

\subsection{Features}
\label{ssec:feats}

\vspace{-2mm}

The system uses sequences of $32\times32$ spectrogram images as input features, similar to those used in \cite{SmallCNN}.
These spectrograms are constructed by computing 32-dimensional log mel-filterbank energies using a frame step of 10 ms, and stacking them together over 320 ms to form one input image.
An example of these spectrogram sequences is shown in \mbox{Figure \ref{fig:architecture}}.
The choice of log mel-filterbank energies was informed by \cite{features}, where it was shown that mel-scaled short-time Fourier transform (STFT) spectrograms perform best out of many standard features used for CNN-based audio classifiers.

\vspace{-2mm}

\subsection{Parameter selection}
\label{ssec:params}

\vspace{-2mm}

\begin{table*}[htb]
    \footnotesize
    \centering
    \begin{tabular*}{\textwidth}{@{\extracolsep{\fill}}lccccccccccc}
        \toprule
            &Conv1 kernel &Conv1 width    &Conv2 kernel 
        &Conv2 width &Dense width   &LSTM width &No. params &Test acc\\
        \midrule
        CNN-BiLSTM$_\text{best}$     & 5$\times$5	& 32	& 3$\times$3	& 128	& 64	& 128	& 531k	& 0.9181\\
        CNN-BiLSTM$_\text{small}$    & 5$\times$5	& 32	& 3$\times$3	& 32	& 64	& 32	& 109k	& 0.9136\\
        \bottomrule
    \end{tabular*}
    \caption{Chosen parameters for the 9$^{\text{th}}$ outer fold models. The number of parameters and the test accuracy are also shown.}
    \label{tab:fold9}
    \vspace{-1mm}
\end{table*}

\begin{figure}
\begin{minipage}[b]{1.0\linewidth}
  \centering
  \centerline{\includegraphics[width=0.8\linewidth]{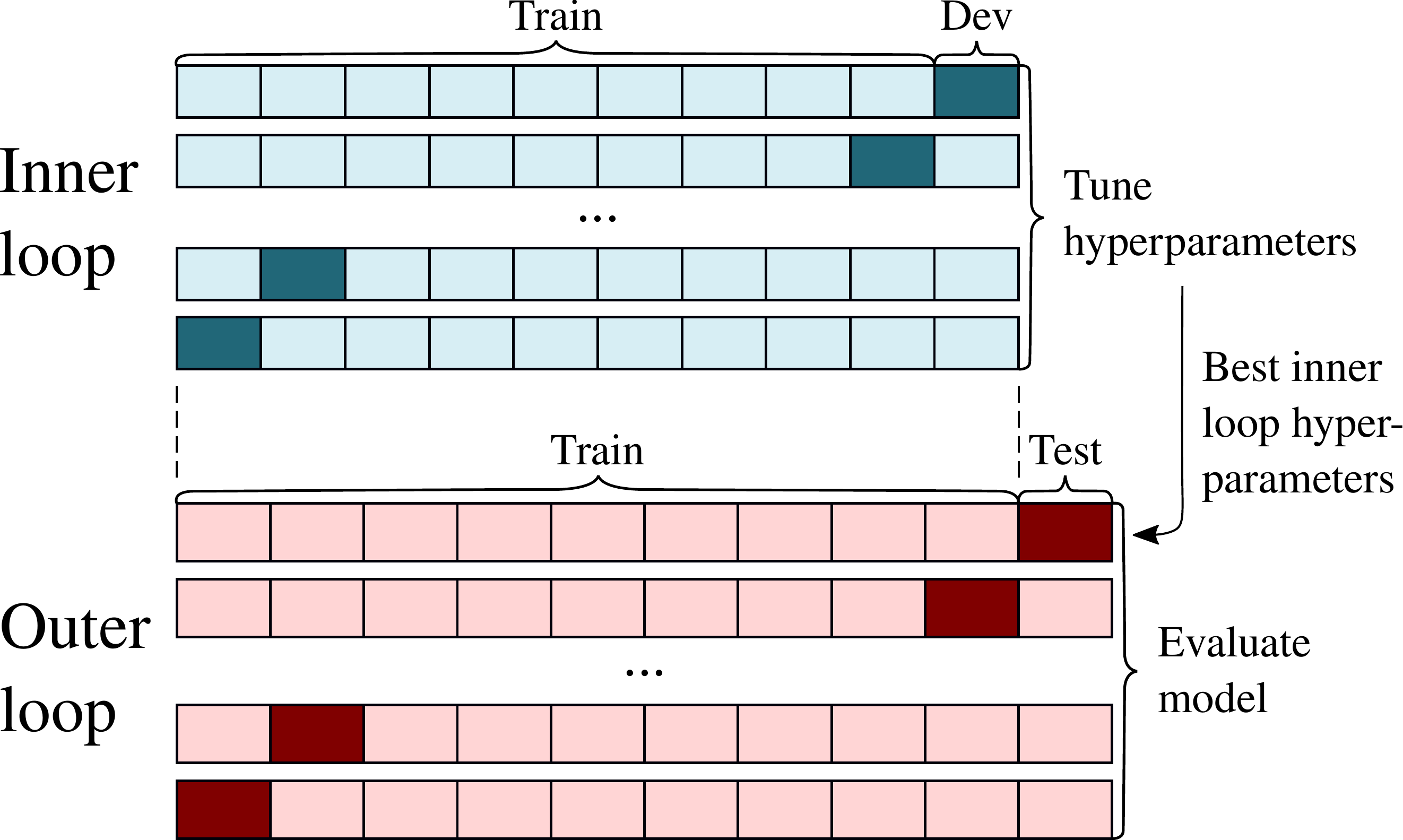}}
\end{minipage}
\vspace{-6mm}
\caption{Nested k-fold cross-validation procedure for hyperparameter tuning and model evaluation.}
\label{fig:kfold}
\vspace{-2mm}
\end{figure}

\begin{figure*}[htb]
  \centering
  \includegraphics[width=1\textwidth]{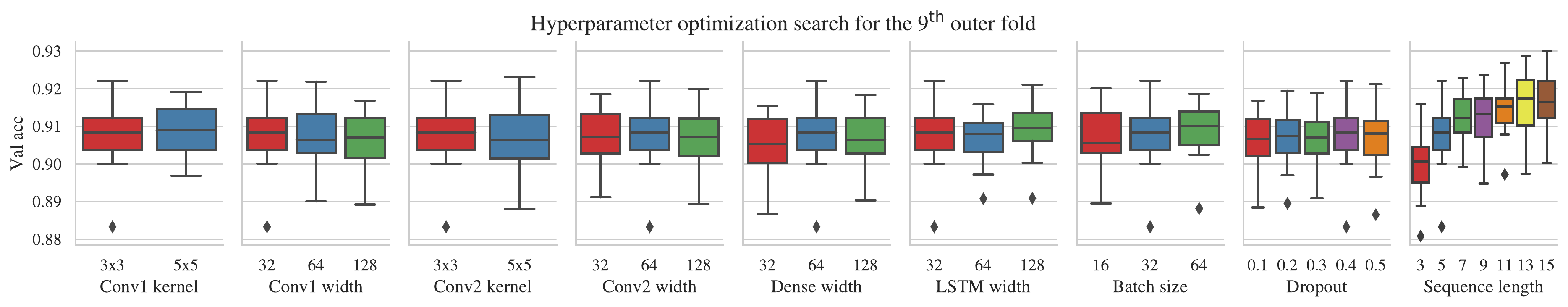}
\vspace{-6mm}
\caption{Boxplots of the inner fold validation accuracy distributions for each hyperparameter in the 9$^{\text{th}}$ outer fold.}
\label{fig:fold9}
\vspace{-3mm}
\end{figure*}

The hyperparameters for this model were optimised using nested 10-fold cross-validation.
As one of the goals for this system is low computational complexity, the relationship between model size and performance is a topic of interest.
Initially a small model was chosen, with fewer than 200k parameters.
Then each parameter was swept across a range of values for each inner training fold to measure the effect on performance.
Different values for kernel filter size, layer widths, dropout, batch size, and training input sequence lengths were evaluated for each inner loop.
The same procedure was used to evaluate the performance of a BiLSTM layer compared to a standard LSTM layer.
The best performing inner loop parameters were used to train models for the respective outer loops, and outer loop models were used for performance assessment.
We refer to the outer loop models trained with best parameters as CNN-BiLSTM$_\text{best}$.
The nested k-fold cross-validation process is shown in \mbox{Figure \ref{fig:kfold}}.

In addition, outer loop models were trained and evaluated for which only the parameters that showed the biggest effects on performance were altered, whilst other parameters were chosen to keep the model small.
We refer to these smaller outer loop models as CNN-BiLSTM$_\text{small}$.
\mbox{Figure \ref{fig:fold9}} and \mbox{Table \ref{tab:fold9}} illustrate the parameter selection process for the 9$^{\text{th}}$ outer fold.
The 9$^{\text{th}}$ fold was chosen for illustration purposes, the results for the other folds are similar.
\mbox{Figure \ref{fig:fold9}} shows the validation accuracy distribution across the inner fold loops for each hyperparameter.
\mbox{Table \ref{tab:fold9}} shows the chosen parameters used for the two outer fold models, informed by the distributions shown in the figure.
Note that only parameters that affect model size are shown in the table.
Optimal batch size, dropout and input sequence length, are used to train both models.
The smaller model has 5 times fewer parameters, while exhibiting a minimal drop in test accuracy.
This process was repeated for all 10 outer folds.
Results are discussed in  \mbox{Section \ref{sec:experiments}}.

\vspace{-3mm}

\section{Experiments and Discussion}
\label{sec:experiments}

\vspace{-2mm}

As described in Section \ref{ssec:params} two collections of outer fold models were trained: CNN-BiLSTM$_\text{best}$ using the best hyperparameters found in the inner fold cross-validation loop, and CNN-BiLSTM$_\text{small}$ using the best hyperparameters that showed a large influence on performance, whilst choosing the other hyperparameters to keep the model size small.
\mbox{Table \ref{tab:param_selection}} shows the resulting model sizes for each outer fold in terms of the number of parameters, and the test accuracies of each model.
We find that while CNN-BiLSTM$_\text{small}$ is on average $<$60\% smaller than CNN-BiLSTM$_\text{best}$, the test accuracy is on average only 0.18\% lower.
In certain folds, such as fold 1, we see that the accuracy of CNN-BiLSTM$_\text{small}$ is actually higher than CNN-BiLSTM$_\text{best}$.
This suggests that the architecture provides stable performance for a variety of hyperparameter configurations.
Furthermore, we note that close-to-optimal performance can be achieved with the substantially smaller models.
This is an important result for ensuring our VAD system is lightweight and computationally efficient.

\begin{figure}
\begin{minipage}[b]{1.0\linewidth}
  \centering
  \centerline{\includegraphics[width=0.9\linewidth]{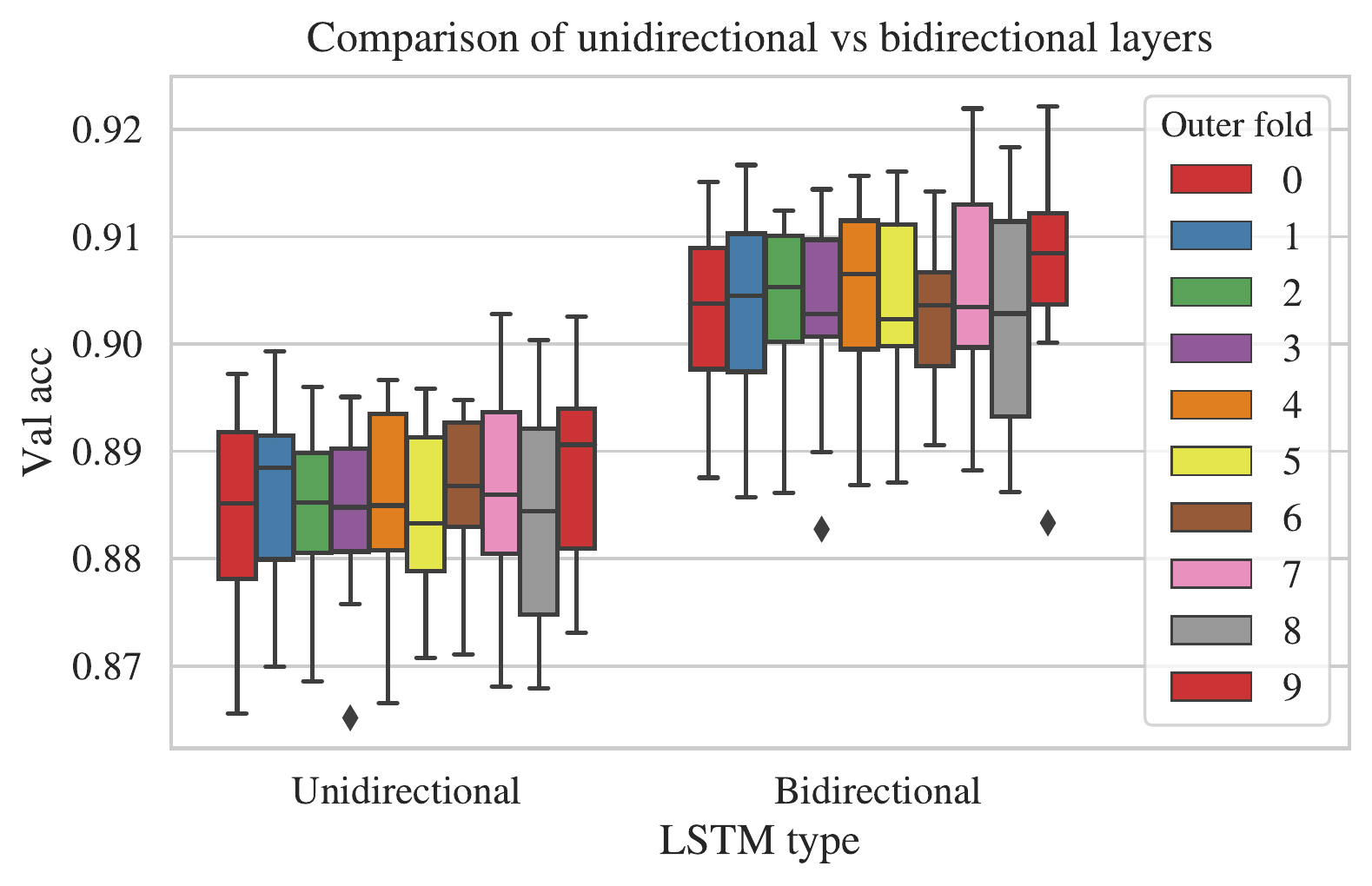}}
\end{minipage}
\vspace{-6mm}
\caption{Validation accuracy across inner fold training loops using a unidirectional LSTM layer vs a BiLSTM layer.}
\label{fig:lstm}
\vspace{-3mm}
\end{figure}

This study also examined the effect of using a unidirectional LSTM layer vs a bidirectional LSTM layer for this architecture.
\mbox{Figure \ref{fig:lstm}} shows the the validation accuracy distributions for each inner fold training loop using both types of LSTM layer.
We find that the BiLSTM layer consistently performs better across all training folds, by $\approx$2\% absolute on average.
A BiLSTM layer would not be a viable option should real-time processing be required.
Fortunately batched processing is acceptable in many applications, including ours.

\begin{table}
    \resizebox{\columnwidth}{!}{
    \centering
    \small
    \begin{tabular}{@{\extracolsep{\fill}}lcccc}
        \toprule
            & \multicolumn{2}{c}{\textbf{CNN-BiLSTM$_\text{best}$}}  & \multicolumn{2}{c}{\textbf{CNN-BiLSTM$_\text{small}$}}\\
        \textbf{Outer fold}  & \textbf{No. params} & \textbf{Test acc} & \textbf{No. params} & \textbf{Test acc} \\ \midrule
        0	& 413k	& 0.9132	& 150k	& 0.9091    \\
        1	& 880k	& 0.8975	& 254k	& 0.9000    \\
        2	& 267k	& 0.9397	& 150k	& 0.9362    \\
        3	& 715k	& 0.9061	& 128k	& 0.9069    \\
        4	& 419k	& 0.9228	& 254k	& 0.9230    \\
        5	& 287k	& 0.9170	& 150k	& 0.9084    \\
        6	& 573k	& 0.9213	& 254k	& 0.9198    \\
        7	& 355k	& 0.9146	& 217k	& 0.9132    \\
        8	& 715k	& 0.9263	& 254k	& 0.9283    \\
        9	& 531k	& 0.9181	& 109k	& 0.9136    \\ \midrule
        \textbf{Mean}  & 516k	& 0.9177	& 192k	& 0.9159    \\
        \textbf{Std dev}  & 204k	& 0.01142	& 60.0k	& 0.01096   \\ 
        \bottomrule
    \end{tabular}
    }
    \caption{Comparison of best and compact outer fold VADs.}
    \label{tab:param_selection}
\end{table}

The AVA-Speech dataset described in Section \ref{sec:data} is accompanied by results from a number of benchmark systems.
First is the widely used WebRTC project VAD \cite{WebRTC}.
The other benchmarks are two CNN-based systems based on the architecture proposed in \cite{BigCNN}.
The smaller of these, \textit{tiny320}, contains three convolutional layers and $<$1M weights, while the other, \textit{resnet960}, is based on the much larger ResNet-50 architecture \cite{resnet50} and has 30M weights.
\mbox{Table \ref{tab:baselines}} reports the results of these baseline systems for the ``CleanSpeech'', ``Speech+Music'' and ``Speech+Noise'' conditions, as well as for all speech across all conditions.
Frame-based true positive rates (TPR) for a fixed false positive rate (FPR) of 0.315, scored over 10ms frames are reported, as described in \cite{AVASpeech}.
The reported values for the CNN-BiLSTM models are the average performance across all outer fold models, with the standard deviations shown in brackets.
For the ``CleanSpeech'' condition both CNN-BiLSTM$_\text{best}$ and CNN-BiLSTM$_\text{small}$ perform on par with the much larger \textit{resnet960} system, which is the best performing baseline.
Under the more difficult conditions ``Speech+Noise'' and ``Speech+Music'' the CNN-BiLSTM systems outperform the baselines by a comfortable margin, particularly for the ``Speech+Music'' condition.
Overall our systems are shown to outperform the best baseline by 5\% absolute.

\begin{table}
    \footnotesize
    \resizebox{\columnwidth}{!}{
    \centering
    \begin{tabular}{@{\extracolsep{\fill}}lcccc}
        \toprule
                        & \multicolumn{4}{c}{\textbf{TPR}}        \\
        \textbf{Model}  & \textbf{Clean} & \textbf{Noise} & \textbf{Music} & \textbf{All} \\ \midrule
        RTCvad          & 0.786     & 0.706	    & 0.733	    & 0.722     \\
        tiny320         & 0.965	    & 0.826	    & 0.623	    & 0.810     \\
        resnet960       & \textbf{0.992}     & 0.944	    & 0.787	    & 0.917     \\ \midrule
        CNN-BiLSTM$_\text{best}$  & \begin{tabular}[c]{@{}c@{}}\textbf{0.992}\\ \textbf{(0.0021)}\end{tabular}
        & \begin{tabular}[c]{@{}c@{}}\textbf{0.961}\\ \textbf{(0.0153)}\end{tabular}
        & \begin{tabular}[c]{@{}c@{}}\textbf{0.950}\\ \textbf{(0.0264)}\end{tabular}
        & \begin{tabular}[c]{@{}c@{}}\textbf{0.968}\\ \textbf{(0.0107)}\end{tabular} \\ 
        CNN-BiLSTM$_\text{small}$  & \begin{tabular}[c]{@{}c@{}}\textbf{0.992}\\ \textbf{(0.0023)}\end{tabular}
        & \begin{tabular}[c]{@{}c@{}}0.960\\ (0.0136)\end{tabular}
        & \begin{tabular}[c]{@{}c@{}}0.948\\ (0.0293)\end{tabular}
        & \begin{tabular}[c]{@{}c@{}}0.967\\ (0.0102)\end{tabular} \\
        \bottomrule
    \end{tabular}
    }
    \caption{The TPR reported at a FPR of 0.315 for various VAD systems tested on AVA-Speech. The CNN-BiLSTM values reported are the means of the outer fold cross-validation results, with the standard deviations shown in brackets.}
    \label{tab:baselines}
    \vspace{-4mm}
\end{table}

\begin{figure}[htb]
\begin{minipage}[b]{1.0\linewidth}
  \centering
  \centerline{\includegraphics[width=0.92\linewidth]{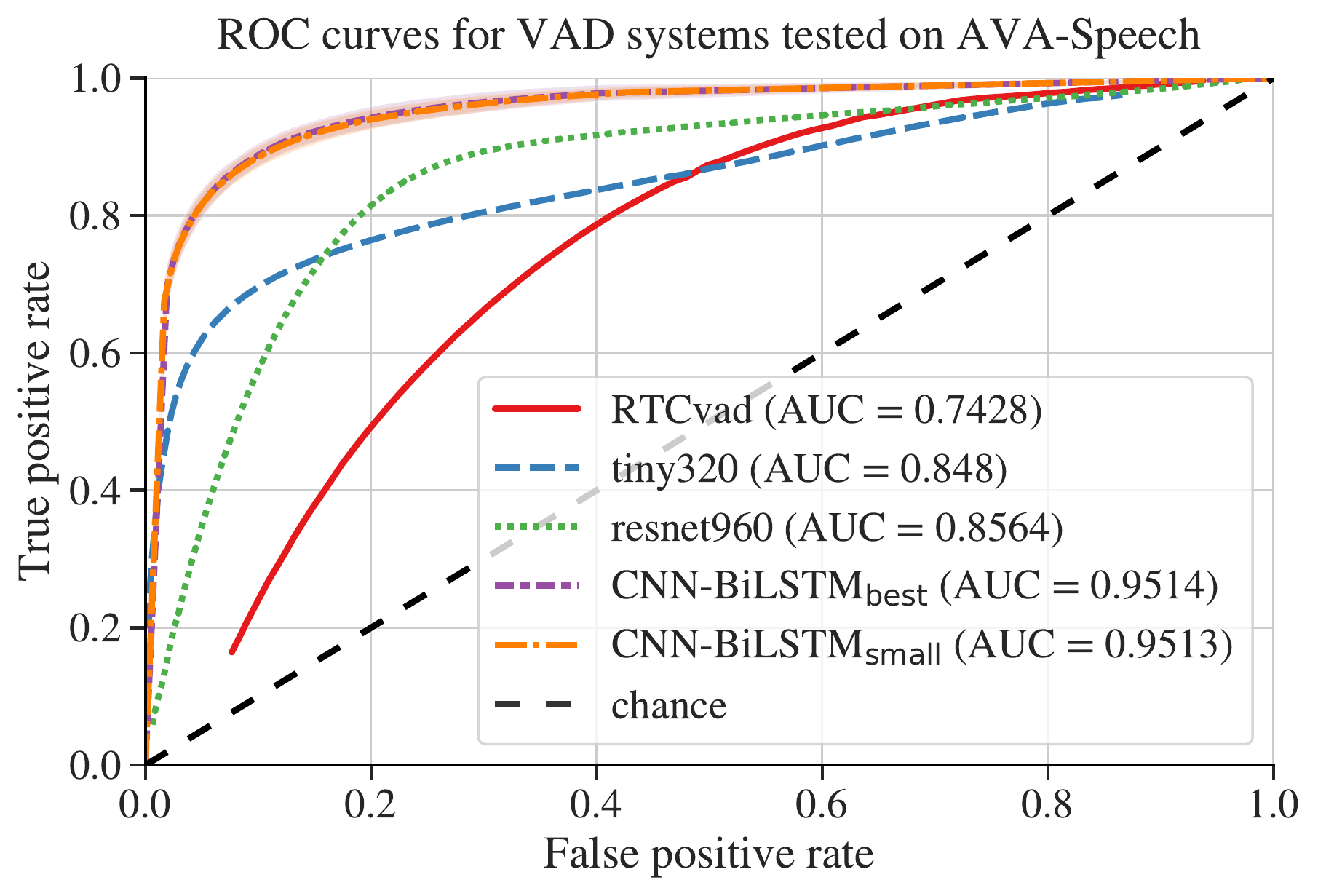}}
\end{minipage}
\vspace{-6mm}
\caption{ROC curves for VADs tested on AVA-Speech. The CNN-BiLSTM curves indicate mean performance over the outer folds, with the standard deviations shown as shaded error regions. The area under the curve (AUC) is also shown.}
\label{fig:roc}
\vspace{-4mm}
\end{figure}

We also again see in \mbox{Table \ref{tab:baselines}} that, despite the much smaller model sizes, CNN-BiLSTM$_\text{small}$ performs on par with CNN-BiLSTM$_\text{best}$.
The larger models only slightly outperform the smaller models for the ``Speech+Noise'' and ``Speech+Music'' conditions.
This is likely due to the capacity of larger networks to more effectively model speech under difficult environmental conditions.
However, the difference is very small indeed.

\mbox{Figure \ref{fig:roc}} shows the receiver operating characteristic (ROC) curves for all systems tested on AVA-Speech.
The curves shown are for all speech conditions, and the results reported for the CNN-BiLSTM systems indicate mean performance over the outer folds, with standard deviations shown as shaded error regions.
We see that the curves of the two CNN-BiLSTM models coincide very closely and that the compact models perform on par with their larger counterparts.
We also note that the standard deviations are very small, indicating stable performance across all outer folds.
Finally, we see that the CNN-BiLSTM systems we propose outperform all baselines across all operating points.

\vspace{-2mm}

\section{Conclusions}
\label{sec:conclusions}

\vspace{-2mm}

This study has introduced a new CNN-BiLSTM model for VAD.
We show that the CNN-BiLSTM architecture not only provides stable performance across a number of hyperparameter configurations, but also that it provides optimal performance with small network sizes.
The impact of using a \mbox{BiLSTM} layer rather than a unidirectional layer has also been explored, and found to be important in achieving optimal performance.
Finally, our CNN-BiLSTM systems were found to outperform previous baseline systems, including a much larger ResNet-based system.
We conclude that this architecture is well-suited to the problem of VAD.
It delivers state-of-the-art performance in difficult \textit{in-the-wild} conditions, whilst remaining lightweight and efficient enough for practical use in resource constrained settings.


\vspace{-2mm}

\section{Acknowledgements}
\label{sec:ack}

\vspace{-2mm}

We would like to thank the South African Centre for High Performance Computing (CHPC) for providing computational resources on their Lengau cluster for this research.


\bibliographystyle{IEEEbib}
\bibliography{refs}

\begin{thebibliography}{10}

\bibitem{TASI}
K.~{Bullington} and J.~M. {Fraser},
\newblock ``Engineering aspects of {TASI},''
\newblock {\em The Bell System Technical Journal}, vol. 38, no. 2, pp.
  353--364, March 1959.

\bibitem{G.729}
``{A silence compression scheme for ITU-T G.729 optimized for terminals
  conforming to ITU-T V.70},''
\newblock Standard ITU-T G.729 Annex B, International Telecommunication Union,
  Geneva, 1996.

\bibitem{GSM06.32}
``{Digital Cellular Telecommunications System (Phase 2+); Full Rate Speech;
  Voice Activity Detector (VAD) for full rate speech traffic channels},''
\newblock Standard GSM 06.32 version 8.0.1, European Telecommunications
  Standards Institute, Valbonne, 1999.

\bibitem{Sohn}
J.~{Sohn}, N.~S. {Kim}, and W.~{Sung},
\newblock ``A statistical model-based voice activity detection,''
\newblock {\em IEEE Signal Processing Letters}, vol. 6, no. 1, pp. 1--3,
  January 1999.

\bibitem{KLTLaplaceGauss}
S.~{Gazor} and W.~{Zhang},
\newblock ``A soft voice activity detector based on a laplacian-gaussian
  model,''
\newblock {\em IEEE Transactions on Speech and Audio Processing}, vol. 11, no.
  5, pp. 498--505, September 2003.

\bibitem{Gamma}
J.~W. {Shin}, J.~{-H} {Chang}, H.~S. {Yun}, and N.~S. {Kim},
\newblock ``Voice activity detection based on generalized gamma distribution,''
\newblock in {\em Proc. ICASSP}, Philadelphia, USA, 2005.

\bibitem{Multiple_models}
J.~{-H}. {Chang}, N.~S. {Kim}, and S.~K. {Mitra},
\newblock ``Voice activity detection based on multiple statistical models,''
\newblock {\em IEEE Transactions on Signal Processing}, vol. 54, no. 6, pp.
  1965--1976, June 2006.

\bibitem{MO_LRT}
J.~{Ramirez}, J.~C. {Segura}, C.~{Benitez}, L.~{Garcia}, and A.~{Rubio},
\newblock ``Statistical voice activity detection using a multiple observation
  likelihood ratio test,''
\newblock {\em IEEE Signal Processing Letters}, vol. 12, no. 10, pp. 689--692,
  September 2005.

\bibitem{SMV1}
D.~{Enqing}, L~{Guizhong}, Z~{Yatong}, and Z~{Xiaodi},
\newblock ``Applying support vector machines to voice activity detection,''
\newblock in {\em Proc. International Conference on Signal Processing},
  Beijing, China, 2002.

\bibitem{SVM2}
J.~Ramirez, P.~Yelamos, J.~Manuel Gorriz, J.~C. Segura, and L.~Garcia,
\newblock ``Speech/non-speech discrimination combining advanced feature
  extraction and svm learning,''
\newblock in {\em Proc. Interspeech}, Pittsburgh, USA, 2006.

\bibitem{SVM3}
Q.~{Jo}, J.~{-H}. {Chang}, J.~W. {Shin}, and N.~S. {Kim},
\newblock ``Statistical model-based voice activity detection using support
  vector machine,''
\newblock {\em IET Signal Processing}, vol. 3, no. 3, pp. 205--210, May 2009.

\bibitem{SVM4}
J.~{Wu} and X.~{Zhang},
\newblock ``Efficient multiple kernel support vector machine based voice
  activity detection,''
\newblock {\em IEEE Signal Processing Letters}, vol. 18, no. 8, pp. 466--469,
  August 2011.

\bibitem{DBN_VAD}
X.~{Zhang} and J.~{Wu},
\newblock ``Deep belief networks based voice activity detection,''
\newblock {\em IEEE Transactions on Audio, Speech, and Language Processing},
  vol. 21, no. 4, pp. 697--710, April 2013.

\bibitem{DNN_VAD1}
N.~Ryant, M.~Liberman, and J.~Yuan,
\newblock ``Speech activity detection on youtube using deep neural networks,''
\newblock in {\em Proc. Interspeech}, Lyon, France, 2013.

\bibitem{RNN_VAD}
T.~{Hughes} and K.~{Mierle},
\newblock ``Recurrent neural networks for voice activity detection,''
\newblock in {\em Proc. ICASSP}, Vancouver, Canada, 2013.

\bibitem{DNN_VAD2}
X.~{Zhang} and D.~{Wang},
\newblock ``Boosting contextual information for deep neural network based voice
  activity detection,''
\newblock {\em IEEE/ACM Transactions on Audio, Speech, and Language
  Processing}, vol. 24, no. 2, pp. 252--264, February 2016.

\bibitem{Biswas2019}
A.~Biswas, R.~Menon, E.~van~der Westhuizen, and T.~R. Niesler,
\newblock ``Improved low-resource somali speech recognition by semi-supervised
  acoustic and language model training,''
\newblock in {\em Proc. Interspeech}, Graz, Austria, 2019.

\bibitem{wilkinson}
N.~Wilkinson, A.~Biswas, E.~Yilmaz, F.~De~Wet, E.~van~der Westhuizen, and T.~R.
  Niesler,
\newblock ``Semi-supervised acoustic modelling for five-lingual code-switched
  {ASR} using automatically-segmented soap opera speech,''
\newblock in {\em Proc. 1st Joint SLTU and CCURL Workshop}, Marseille, France,
  2020.

\bibitem{BigCNN}
S.~{Hershey}, S.~{Chaudhuri}, D.~P.~W. {Ellis}, J.~F. {Gemmeke}, A.~{Jansen},
  R.~C. {Moore}, M.~{Plakal}, D.~{Platt}, R.~A. {Saurous}, B.~{Seybold},
  M.~{Slaney}, R.~J. {Weiss}, and K.~{Wilson},
\newblock ``{CNN} architectures for large-scale audio classification,''
\newblock in {\em Proc. ICASSP}, New Orleans, USA, 2017.

\bibitem{SmallCNN}
A.~{Sehgal} and N.~{Kehtarnavaz},
\newblock ``A convolutional neural network smartphone app for real-time voice
  activity detection,''
\newblock {\em IEEE Access}, vol. 6, pp. 9017--9026, February 2018.

\bibitem{AVASpeech}
S.~Chaudhuri, J.~Roth, D.~P.~W. Ellis, A.~C. Gallagher, L.~Kaver, R.~Marvin,
  C.~Pantofaru, N.~Reale, L.~G. Reid, K.~W. Wilson, and Z.~Xi,
\newblock ``{AVA-Speech}: A densely labeled dataset of speech activity in
  movies,''
\newblock in {\em Proc. Interspeech}, Graz, Austria, 2018.

\bibitem{CLDNN}
T.~N. {Sainath}, O.~{Vinyals}, A.~{Senior}, and H.~{Sak},
\newblock ``Convolutional, long short-term memory, fully connected deep neural
  networks,''
\newblock in {\em Proc. ICASSP}, Brisbane, Australia, 2015.

\bibitem{CLDNNVAD}
R.~Zazo, T.~N. Sainath, G.~Simko, and C.~Parada,
\newblock ``Feature learning with raw-waveform {CLDNN}s for voice activity
  detection,''
\newblock in {\em Proc. Interspeech}, San Francisco, USA, 2016.

\bibitem{Adam}
D.~P. Kingma and J.~Ba,
\newblock ``Adam: A method for stochastic optimization,''
\newblock in {\em Proc. International Conference on Learning Representations},
  San Diego, USA, 2015.

\bibitem{features}
M.~Huzaifah,
\newblock ``Comparison of time-frequency representations for environmental
  sound classification using convolutional neural networks,'' June 2017,
\newblock [Online]. Available: \url{https://arxiv.org/pdf/1706.07156}.

\bibitem{WebRTC}
WebRTC.org,
\newblock ``The {WebRTC} project,'' 2011,
\newblock [Online]. Available: \url{https://webrtc.org}.

\bibitem{resnet50}
K.~{He}, X.~{Zhang}, S.~{Ren}, and J.~{Sun},
\newblock ``Deep residual learning for image recognition,''
\newblock in {\em Proc. CVPR}, Las Vegas, USA, 2016.

\end{thebibliography}

\end{document}